\documentclass[twocolumn, showpacs,preprintnumbers,amsmath,amssymb]{revtex4} 
\usepackage{}

\usepackage{graphicx}
\usepackage{dcolumn}
\usepackage{bm}

\begin{document}

\title{Cascading Walks Model for Human Mobility Patterns}

\author{Xiao-Pu Han$^{1}$}\email{xp@hznu.edu.cn}
\author{Xiang-Wen Wang$^{2,3}$}
\author{Xiao-Yong Yan$^{4,5}$}
\author{Bing-Hong Wang$^{3}$}

\affiliation{$^{1}$Institute of Information Economy and Alibaba Business College, Hangzhou Normal University, Hangzhou 310036, China\\
$^{2}$Department of Physics, Virginia Polytechnic Institute and State University, Blacksburg, Virginia 24061-0435, USA\\
$^{3}$Department of Modern Physics, University of Science and Technology of China, Hefei 230026, China\\
$^{4}$Research Center for Complex System Science, Normal University of Beijing, Beijing 100875, China \\
$^{5}$Web Sciences Center, University of Electronic Science and Technology of China, Chengdu 610051, China}

\date{\today}

\begin{abstract}
Uncovering the mechanism behind the scaling law in human trajectories is of fundamental significance in understanding many
spatio-temporal phenomena. In combination of the exploration and the preferential returns, we propose a simple dynamical
model mainly based on the cascading processes to capture the human mobility patterns. By the
numerical simulations and analytical studies, we show more than five statistical characters that are well consistent with the empirical observations, including several type of scaling anomalies, and the ultraslow diffusion property, implying the cascading processes associated with the other two mechanisms are indeed a key in the understanding of human mobility activities. Moreover, both of the diverse individual mobility and aggregated scaling move-lengths, bridging the micro and macro patterns in human mobility.
Our model provides deeper understandings on the emergence of human mobility patterns.

\end{abstract}


\maketitle

\section{Introduction}


In recent years, benefited from the development of location tracking technologies, the studies on the statistics of human daily mobility rise to be a new research hot point. Researchers have analyzed varieties of data sets reflecting human mobility (e.g. dollar-bill tracking, mobile phone roaming, global positioning system, recording of taxi and bus passengers, et al).
These empirical studies show two major quantitative features of human mobility patterns: i). the scaling anomalies in several statistics, which mainly includes the aggregated jump-size (or say move-length or displacement) of each travel \cite{bro,gon,Rhee,Lee,Jiang2009, Song2010b,Jia,Szell}, waiting time \cite{gon, Song2010b}, visitation frequency of each distinct locations \cite{gon, Song2010b}, and the growth of the number of visited locations \cite{gon, Song2010b}. ii). Ultraslow diffusion property, which is observed in the slowly growth of the mean square displacement (MSD) and radius of gyration \cite{bro,gon,Rhee,Song2010b}. Beyond that, several other anomalies, such as high regularity and predictability \cite{Song2010} and the limitation of traffic system \cite{Jiang2009}, have also been observed in human mobility.
These features are not only against the traditional view of human mobility based on random walks but also sharply different with the purely L\'evy-flight nature. Consequently, researchers have to review the discussions on many social and economic dynamics that are deeply affected by human mobility, such as the spreading of pathogens and information \cite{Belik,Balcan,WangL,Ni,Wang2009,Zhao}, the planning of city and traffic systems \cite{Horner}, and so on.

Moreover, with the progress on this issue, even some recently formed consensus are also argued by the new findings. For instance, the aggregated distribution of long-range move-length shows widely scaling properties, however, this conclusion can not be extend to be the universal law of human mobility \cite{Petro}, because human individuals' movements show high diversity \cite{Yan2} and the pattern of urban trips usually appear to be exponential-like \cite{Baz1,Liang,Noulas,Peng}.


A key issue is to understand the origin of these properties. Recent studies have proposed several approaches and dynamical models to describe it.
The simplest type is the descriptive model, including the continuous-time random-walk (CTRW), which can generate power-law-like displacement distribution and show slow diffusion \cite{bro}, and the self-similar least action walk (SLAW) \cite{Lee}.
In the explanation of the scaling anomalies, many possible origins have been proposed. The first type is the aggregated effect on population to mimic the macroscopic features of human mobility. Such as the radiation model that can reproduce much realistic inter-city mobility patterns \cite{Simini}, and the model based on maximum entropy theory under Maxwell-Boltzmann statistics which aggregate the scaling law from diverse individual movements \cite{Yan2}.  In the second type, the spatial heterogeneity of population density and urban environment, was discussed as a factor that contributes to the emergences of human mobility patterns \cite{Noulas,Vene,Baz2}. Beyond these qualities, recently, the hierarchical nature in human mobility is empirically observed \cite{Jia} and is considered as one of the origin of the scaling move-length \cite{Han, Jia}.
For the ultraslow diffusion, the model reported in Ref. \cite{Yan1} indicates that it seem to be the result of the high-frequency trips between homes and working locations.
Recently, the model proposed by Song et al \cite{Song2010b} attract much attention. It considers two generic mechanisms: the exploration of new locations and the preferential return of visited locations.
In the combination of some priori results obtained in empirical analysis, this model provide several statistical patterns agreeing with the empirical data,
indicating that the two mechanisms play fateful role in human mobility. Unfortunately, due to the dependence on the priori empirical results, this model does not give a sufficient explanation on the characteristic key of human mobility patterns: the scaling law on jump-size.
Overall, a model that catches the key in human mobility and reproduces most of empirical observations is still lacking.

In the present paper, based on the cascading process and combination of the localized exploration and the preferential returns, we propose an agent-based model to explain the underlying mechanisms in human daily travels. We surprisingly find that, this minimum model can quantitatively well fit almost all the empirical findings and cover most of the results of previous modeling studies, implying the cascade-like process is indeed important in the understanding of human mobility patterns.


\section{The model}

\begin{figure}
\includegraphics[width=8.5cm]{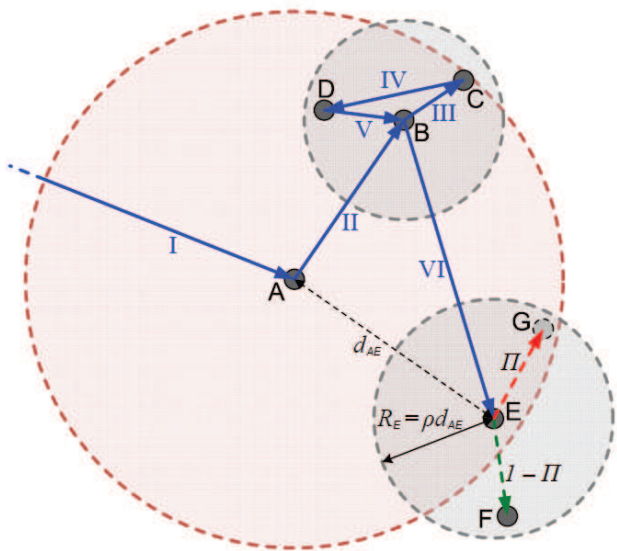}\\
\caption{(Color online) Illustration of the cascading walks process in the model. It shows several consecutive movements of the walker.}\label{Model}
\end{figure}

\begin{figure*}
\includegraphics[width=17cm]{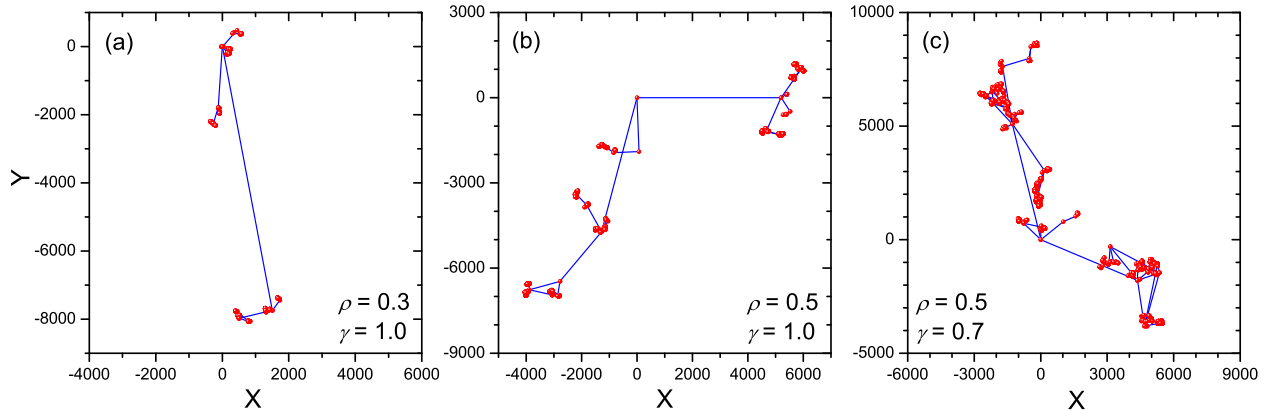}\\
\caption{(Color online) Illustration of the trajectories of 5000 consecutive travels generated by the model with three typical parameter settings on $\rho$ and $\gamma$. }\label{Traj}
\end{figure*}


One noticeable feature of our daily long-range trips and travels is that, we usually prefer to take some short-range movements around the aiming locations rather than return homes directly. For example, when we travel from our hometown to another city, generally speaking, we need to find a hotel, then we will take some activities for the travel (e.g., working, meeting, or tours) at the locations nearby the city, and at last we return to hometown after all objectives are completed. Each things that we do in the travel will create one or more short-range movements in the field of the city, that is to say, the long-range travel from our hometown to the city activates series shorter movements in the aiming city. Similarly, such activation also can be observed in many middle-range or short-range movements, for instance, the commuter's trips vs. the movements in workplaces, and the trips from home to marketplace vs. the strolls during shopping. Associating with the activation by long-range movements, an explicit feature of human mobility emerges: series activating processes act most of transitions from relatively longer travels to shorter trips, and each of them can initiate series shorter trips around the end location of the longer travel. These serial initiations of movements can be called ``Cascading Walks Process".

This cascading walking process in human mobility have been observed by the recent empirical study based on the database of GPS carriers \cite{XWWang}, in which the evidences of cascading walks process that the positive correlation between the displacements of two consecutive movements, positively relates to the scaling property of human mobility patterns, implying the cascading walks process might play an important role in the emergence of human scaling mobility patterns. Also, previous studies in the temporal patterns of human activities indicated a similar cascading-like process would be relevant to the emergence of burst \cite{Malmgren}, as well as the long-term correlations (can be created by the cascading processes) \cite{Rybski2009,Rybski2012}.
Overall, combining all the daily experiences and empirical studies above, the cascading walks process is regarded as the basic mechanism of our modeling work that will be elucidated later.

Initiations of series of shorter movements after a longer travel is the basic dynamics of cascading walks process, and thereby in the present model, the longer travel and initiated shorter movements can be distinguished into two layers: the longer travel is in a higher layer, and the shorter movements are in the lower one, and the corresponding end locations also can be set into different layers. An illustration for the algorithm of the cascading walks in the present model is shown in Fig. \ref{Model}. Associating with Fig. \ref{Model}, the detailed rules will be shown below.

i). The localized cascading process. It's the basic rule of the present model. We assume that the walker moves into an $i$-th layer location A (the movement I in Fig. \ref{Model}). The long-range movement I is also in the $i$-th layer and initiates several ($i+1$)-th layer shorter movements in the field of A (the black dashed circle), which are II (A $\rightarrow$ B) and VI (B $\rightarrow$ E), as shown in Fig. \ref{Model}. The two ($i+1$)-th layer locations B and E are randomly chosen in the field of A. The movements II and VI respectively initiate several ($i+2$)-th layer short trips in the field of B and E (the two gray dashed circles, and they are called the sub-field of A), for instance, the movements III, IV and V in the field of B. These ($i+2$)-th layer trips also could be initiate ($i+3$)-th layer trips (it's not shown in Fig. \ref{Model}). The radius $R's$ of the field of B/E, are respectively proportional to the distances from A to B/E: $R_B = \rho d_{AB}$ for location B, as well as $R_E = \rho d_{AE}$ for E, where $\rho$ is a parameter satisfying $ 0 <\rho < 1$. Here a part of the field of B/E outside the field of A is allowed.

ii). The staying time is correlated to the distance.
For the location B, say, the total time of all these lower-layer short movements which are directly or indirectly initiated by the movement II is $\tau_B$, namely $\tau_B$ is the time length from the movement III to V, or say, the staying time of B. We assume the staying time $\tau_B$ are positively correlated to the distance from the up-layer location A to location B,
\begin{equation}
\tau_B = f_P(d_{AB}^{\gamma})
\end{equation}
which means $\tau_B$ is a random number obeys a Poisson distribution $f_P$ with average value $d_{AB}^{\gamma}$, as well as $\tau_E = f_t(d_{AE}^{\gamma})$ for the location E, and similarly for others, for instance, $\tau_C = f_P(d_{BC}^{\gamma})$, where $\gamma$ is another main parameter of the model, denoting how the staying time correlates to the distance to the up-layer location.
In the simulations, when the staying time will be exhausted, we set the walker must return to the center location of the current field to prepare the departure to other field, for instance, when $t = t_B + \tau_B -1$, the walker moving in the field of B must return to B, no matter whether the staying time of the current field is exhausted or not.

iii). Exploration and preferential return. During the term that the walker moves into a field of a location (the ($i+1$)-th layer location E, say), as shown in Fig. 1, for each initiation of a new ($i+2$)-th layer movements, with probability
\begin{equation}
\Pi = \dfrac{\lambda }{\lambda +\Sigma_j k_j}
\end{equation}
the walker can visit a new ($i+2$)-th layer location (G, say) from its present location, where G is randomly chosen from the field of E,  and $\Sigma_j k_j$ denotes the total number of visitations of all the ($i+2$)-th layer locations in the current ($i+1$)-th layer field except the current location. And with probability $1-\Pi$, the walker can return to a visited ($i+2$)-th layer location (F, say),  chosen with a probability that is proportional to its visitation number $k$. In other words, in this field and this layer, the location visited more frequently are more possible to be visited again. Note that, the ($i+1$)-th layer location E is also treated as one of these ($i+2$)-th layer locations in the field. If E is not the current location, as well as A for the the ($i+1$)-th layer locations in its field. The walks in other layers, e.g. A $\rightarrow$ B and B $\rightarrow$ E, also obey the above algorithms. Comparing with the algorithm of Song's model \cite{Song2010b}, the exploration and preferential return are localized in the current field and current layer.

\begin{figure*}
\includegraphics[width=16cm]{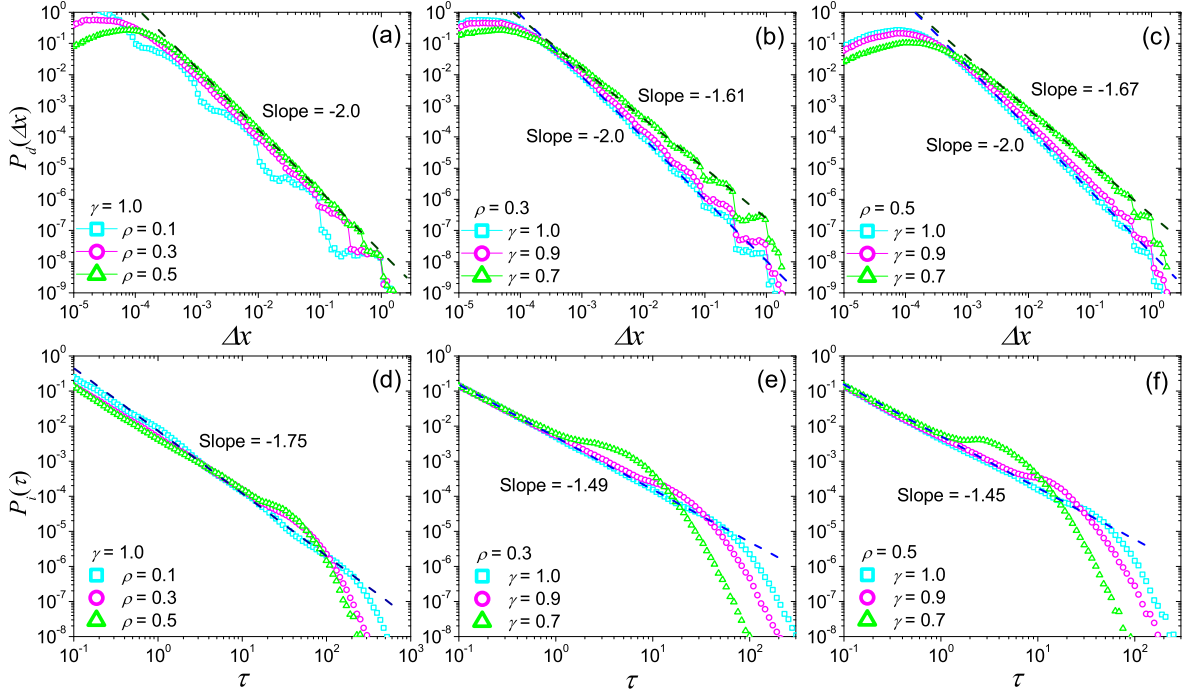}\\
\caption{(Color online) (a), (b) and (c) respectively are the displacement distributions generated by the present model with different parameters $\rho$ and $\gamma$, as well as (d), (e) and (f) for the inter-visit time distributions. The result is averaged by 1000 independent runs. The dark blue dashed lines and dark green dashed lines respectively corresponds to the fitting lines of the data signed by light blue squares and light green rectangles.}\label{Px}
\end{figure*}

\begin{figure}
\includegraphics[width=8.5cm]{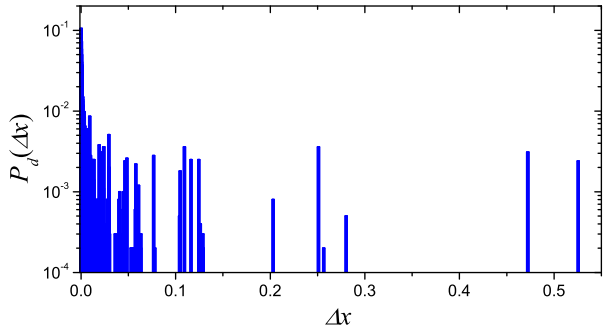}\\
\caption{(Color online) The displacement distribution of an agent, which corresponds to the simulation shown in Fig. \ref{Traj}(c).}\label{Singleuser}
\end{figure}

\section{Simulations}

At the initial time $t = 0$ of simulations, the walker starts its travel from the initial location with coordinate (0, 0) at the top layer. The radius of the field of the initial location, which also is the maximum radius of all the fields, is fixed to be $R_0 = 10^4$. The value of $\lambda$ is fixed to be 0.5. All the statistical results that will be discussed below are obtained from the simulations with $5\times 10^4$ time steps after an initialization with same time steps. In the following discussions, the relative displacement $\Delta x_t = \frac{|\vec{X}_{t+1} - \vec{X}_t|}{R_0}$ are used to replace the absolute value of displacement, here $\vec{X}_{t}$ and $\vec{X}_{t+1}$ are respectively the coordinate of the agent at time step $t$ and $t+1$.

The patterns of trajectories generated by the present model are very similar to our intuitive experience of daily life. Fig. \ref{Traj} shows the typical trajectory patterns under different parameter settings. The trajectories in different scales have significant self-similar property, and large $\rho$ and small $\gamma$ can generate more complicated trajectory patterns.
Most importantly, we surprisingly find out that the present model can create most of the properties observed in empirical studies, including both the scaling anomalies and ultraslow diffusions. In the following, we will discuss them one by one.

\begin{figure*}
\includegraphics[width=16cm]{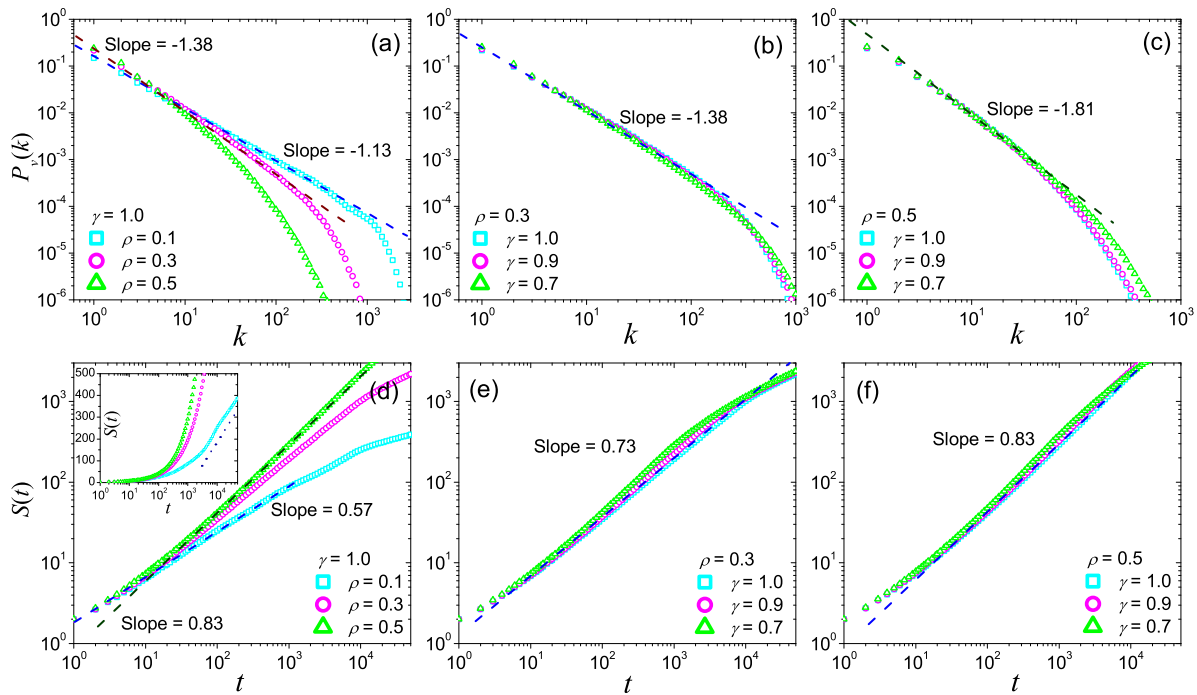}\\
\caption{(Color online) (a), (b) and (c) respectively are the visiting frequency distributions generated by the present model for different parameters $\rho$ and $\gamma$, as well as (d), (e) and (f) with the growth of the number of visited locations. The result is averaged by 1000 independent runs. The dark blue dashed lines and dark green dashed lines respectively corresponds to the fitting lines of the data signed by light blue squares and light green rectangles, and the blue dashed line in the inset of (d) shows the logarithmical section.}\label{Visit}
\end{figure*}

i) Displacement distributions.

The scaling statistics of move-lengths is one of the most noticeable differences between real-world human mobility patterns and the random walk nature, which has also been observed in some animals movements \cite{bart1,Ramos, Sims}. Empirical studies have provided lots of evidences to support a widely accepted viewpoint that the displacement distributions of human travels at the population level are power-law-like \cite{bro,gon}, no matter the arguments at the individual level.
Now, understanding the origin of scaling displacement distribution becomes an important task to modeling researches. Numerical simulations indicate the displacement distribution with a power-law tail can be emerged in all parameter space of our model (Fig. \ref{Px}(a), (b) and (c)). The exponent $\alpha$ of power-law tail is mainly affected by $\gamma$, and is insensitive to other parameters. The value of $\alpha$ is $2.0$ when $\gamma = 1.0$, and reduces along with the reduction of $\gamma$ (Fig. \ref{Px} (b) and (c)), covering the range of many empirical studies. When the value of $\rho$ is small, even though the fluctuations on the curve of displacement distribution are obvious, it does not damage the scaling property in the tail (Fig. \ref{Px} (a), (b) and (c)).
The analytical result (see {\it Appendix A}) indicates that, the exponent $\alpha$ only depends on $\gamma$: $\alpha = 1 + \gamma$, and the scaling displacement distribution essentially emergences from the localized cascading process, which determines the hierarchically organization of locations.

Note that, the results in Fig. \ref{Px} are obtained by an average of large number of simulations. Each run can be regard as a simulation for an agent, accordingly $P_d(\Delta x)$ in Fig. \ref{Px}(a-c) actually shows the aggregated patterns of displacement distributions in global level, which corresponds to most of empirical studies. What does the real-world travel patterns at the individual level look like is still in controversy. Recent studies reported some evidences that, human mobility shows rich diversity at the individual level, and some do not obey the scaling displacement distribution \cite{Yan1,Yan2}. The diversity and non-scaling properties in the individual-level mobility are created by the present model. We plot the displacement distributions generated by a single run and find most of them have not clear power-law tail. Fig. \ref{Singleuser} shows a typical example of the displacement distribution at the individual level, in which the tail is multi-peak-like due to the repeats on the agents' trajectories, generally in agreement with the empirical findings reported by Ref. \cite{Yan2}.

ii) Inter-visit time distributions.

The inter-visit time of a location is the time interval of two consecutive visits of the walker. Although the direct observations on the inter-visit time distribution of human movements is absent in empirical studies, some evidences still can be obtained indirectly from the empirical analysis of the human behaviors in special site. Such as the power-law-like inter-event time distributions of library borrowing \cite{Vqz2006}, implying the inter-visiting time distribution of many locations could have non-Poisson properties.

Power-law-like inter-visit time distributions are surprisingly emerged in our model with the slope between -1 and -2 (Fig. \ref{Px}(d), (e) and (f)) with different parameter settings. The curves of the inter-visit time distribution $P_i(\tau)$ vary as $\gamma$ decrease, from a power-law function to a bimodal form with a power-law head. The fitting power-law exponent $\zeta$ of the head of $P_i(\tau)$ is close to 1.5 and somewhat negatively correlate to value of $\rho$ (Fig. \ref{Px}(d), (e) and (f)), but is not sensitive to $\gamma$.  For large $\rho$, a tiny peak appears in the tail of $P_i(\tau)$ and moves to the front when $\gamma$ is small.



\begin{figure*}
\includegraphics[width=16cm]{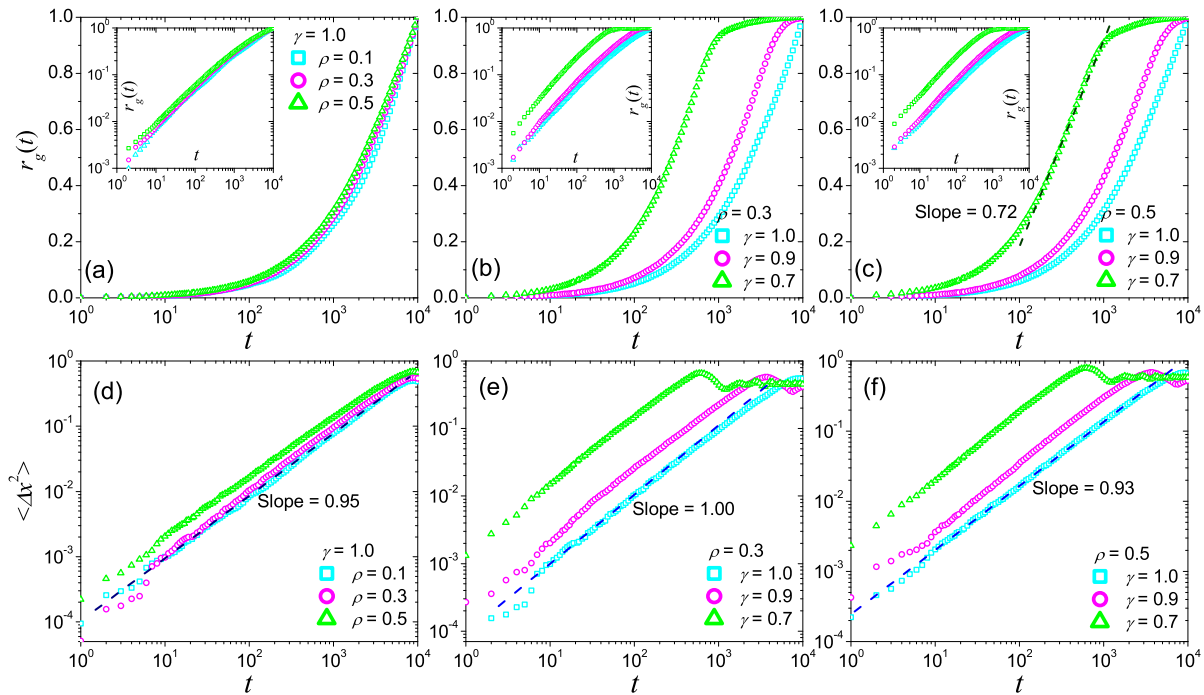}\\
\caption{(Color online) (a), (b) and  (c) respectively are the growth of radius of gyration generated by the present model with different parameters $r$ and $\gamma$, as well as (d), (e) and (f) for the MSD. The result is averaged by 1000 independent runs. The dark blue dashed lines and dark green dashed lines respectively corresponds to the fitting lines of the data signed by light blue squares and light green rectangles. }\label{RgMSD}
\end{figure*}

iii) Visitation frequency and the number of visited locations

Empirical studies of human mobility patterns have indicated that, a few locations, which generally are working places and supermarkets, are visited much more frequently than other places in our daily life.
This heterogeneity is described by the power-law-like visitation frequency distributions in Zipf's ranking plot with the Zipf's exponent $\xi \approx 1.2$ \cite{gon,Song2010b}.
According to the relationship $\xi' = 1+ \xi^{-1}$ between the Zipf's exponent $\xi$ and $\xi'$, where $\xi'$ is the exponent of the correspondingly power-law probability density function (PDF), we can get a PDF exponent $\xi'$ of the visitation frequency distribution about $1.8$.

This scaling visitation frequency is reproduced by the present model. The curves in Fig. \ref{Visit}(a-c) illustrate the power-law-like PDF $P_v(k)$ of visitation frequency for different locations under different parameter settings. $P_v(k)$ is very sensitive to the value of $\rho$: The fitting value of the PDF exponent $\xi'$ reduces from 1.8 to 1.1 along with the growth of $\rho$ from 0.1 to 0.5. The analysis in {\it Appendix B} indicates that $\xi'$ in a extreme case without the cascading effect is $1.0$, implying that these larger values of $\xi'$ possibly originate from the cascading effect.

The growth of the number of visited locations $S(t)$ is generally expected to the scaling form $S(t) \sim t^{\mu}$.  $\mu = 1$ stands for the pure Le\'vy flights, and its value depends on the exponent of the staying time distribution, where staying time is the time that a walker spend at one location in CRTW \cite{Song2010}.
Empirical studies have indicated that, in the real-world $S(t)$ obeys scaling law with an exponent $\mu \approx 0.6$, which is smaller than the prediction in CRTW \cite{gon, Song2010}, indicating an ultraslow growth of the number of visited locations and a growing returning probability to the visited locations. The growth of $S(t)$ in the present model also obeys the power-law function with an exponent $\mu<1$. The visitation frequency is of the similar situation, the exponent is mainly affected by the value of $\rho$, where smaller $\rho$ corresponds to a slower growth of $S(t)$ (small $\mu$), indicating the walker are more possible to return to the visited locations. An extreme situation without the cascading effect shows a logarithmical-growing $S(t)$ ({\it Appendix B}), that is the reason that $S(t)$ of large $\rho$ generates logarithmic sections (Fig. \ref{Visit}(d)). Thus we infer that the cascading effect deeply changes $S(t)$ from the logarithmic form to the power-law growth.



iv) Radius of gyration and MSD.

The properties of human spreading are mainly shown in the growth of radius of gyration $r_g$ and MSD along time.
For the discrete time model, the radius of gyration of a walker in the time period (0, $t$) is defined as:
\begin{equation}
r_g(t) = \sqrt{\frac{1}{t}\sum_{i=1}^{t}(\vec{x}_i - \vec{x}_m)^2},
\end{equation}
where $\vec{x}_i$ represents the position of each visited location, and $\vec{x}_m$ is the center mass of the trajectory in the period \cite{gon}. Empirical studies noted that the growth of human $r_g(t)$ is close to the logarithmic form rather than the power function of CRTW and Le\'vy flight, indicating an ultraslow spread in human long-range travels \cite{gon, Song2010b}.
This abnormal property is also reproduced by the present model. As shown in Fig. \ref{RgMSD}(c), $r_g(t)$ shows a partly logarithmic-like growth when $\gamma<1$, which generally is close to the empirical findings on the mobility of mobile users \cite{gon, Song2010b}.

The MSD $\langle \Delta x^2 \rangle$ at time $t$ is defined as:
\begin{equation}
\langle \Delta x^2(t) \rangle = \langle (\vec{x}_t - \vec{x}_0)^2\rangle
\end{equation}
where $\vec{x}_0$ is the coordinates of initial position and $\langle \cdot \rangle$ means the average of all the time range. Previous studies indicated that, $\langle \Delta x^2(t) \rangle \sim t$ for Brownian random walks and $\langle \Delta x^2(t) \rangle \sim t^{1/(\alpha -1)}$ for pure L\'evy flight, and the MSD of real-world human mobility is generally slower than the prediction of L\'evy flight \cite{Song2010b, Rhee}.

The curves of MSD created for different parameter settings are shown in Fig. \ref{RgMSD}(d-f). The fitting exponents are between in 0.9 and 1.0, which is generally close to the empirical findings based on GPS carriers \cite{Rhee}. For the cases $\gamma<1$, due to the fitting exponent $\alpha$ of the displacement distribution $P_d(\Delta x)$ is less than 1.0, the growth of MSD in our model is slower than the prediction of pure L\'evy flight and more similar to the empirical findings.

These results indicate that the ultraslow diffusions mainly emerge in a condition when $\gamma<1$.  In this situation, the time that the walker moves in the field of lower layer location is relatively short, and the walker therefore has more frequent returns to the upper layer locations and consequently results in the ultraslow diffusions.

\section{discussions}


The basic assumption of the model is the cascading process on both the spatial and temporal correlations, which is in supporting of several recent empirical observations \cite{Jia,XWWang}. For another assumption: the localized preferential returns and explorations, compared with Song et al's model \cite{Song2010b}, its expression is linear, and its effect is localized. Besides these two assumptions, this model do not introduce any other spatio-temporal heterogeneities, implying the environments heterogeneity (e.g. population density, urban environments, et al) \cite{Noulas, Vene} does not play the fateful role in the emergence of the scaling properties in long-range human mobility.

This minimum model surprisingly reproduce many statistics achieving of the empirical data, including the scaling anomalies in move-length, inter-visit time and visitation frequency, the slow growth of the number of visited locations, radius of gyration and MSD, thus it covers most of empirical findings.
For the scaling displacements that is the most noticeable feature in real-world human mobility, the present model successfully creates it without any prior scaling properties. The analytical result of this model indicates that the cascade-like process is the key in the emergence of scaling move-lengths.

Moreover, the present model provides an unified explanation for the differences on empirical findings in different levels and bridges several controversial issues.
First, it shows both the aggregated scaling law and diversity of individual mobility, in agreement with the recent empirical findings crossing the macroscopic and microcosmic domains \cite{Yan2}, indicating a possible mode in the emergence of the aggregated scaling properties of large number of non-scaling individuals.
Secondly, the result that the mobility in single layer do not show the scaling jump-length, is similar to the real-world situations with single type of vehicle (e.g. bus, taxi, et al) \cite{Liang}, successfully explaining both the scaling anomalies on move-length in global level and the homogeneous properties in urban mobility and single-vehicle trips \cite{Noulas, Yan2, Baz1, Baz2, Liang}.
In addition, the scaling visitation frequency in the model implies the existence of dominant trips, in agreement with peoples' daily experiences and empirical findings \cite{gon, Song2010b, Yan2}.

In summary, the present model shows that, only based on the two basic mechanisms, the cascading walks process and the localized preferential returns and explorations, are enough to explain most of the statistics for human mobility in different levels. This conclusion would be very helpful in the understanding of the underlying mechanisms driving human mobility patterns and valuables in the prediction of human daily trips \cite{Domenico} and further applications.





\begin{acknowledgments}
This work is supported by the National Natural Science Foundation of China (11205040, 11105024, 11275186, 70971089), and the Zhejiang Provincial Natural Science Foundation of China under the grant No. LY12A05003, and the research startup fund of Hangzhou Normal University.
\end{acknowledgments}

\appendix

\section{Analytical Results of the displacement distributions}

In the present model, according to the basic rule of the model: $R = \rho d_c$, $\tau = f_p(d_c^{\gamma})$, and assuming $\overline{\Delta x}_1 = 1$ for the first layer movements, the averaged displacement in the $i$-th layer movements and the corresponding staying time can be written immediately:
$\overline{\Delta x}_i = \rho_*^{i-1}$ and $\overline{\tau}_i = \overline{\Delta x}_i^{\gamma} = \rho_*^{(i-1)\gamma}$.
where $\rho_*$ is the ratio of the averaged radius of the $(i+1)$-th layer and $i$-th layer fields: $\rho_* = \overline{R}_{i+1}/\overline{R}_i$. Obviously, $\rho_*=2\rho/3$, since the center of the $(i+1)$-th layer is randomly chosen in the $i$-th layer field.

Due to each $i$th-layer movements can activate several $(i+1)$-th layer trips, averagely, we have
\begin{equation}
\Omega_{i+1} = \Omega_i\tau_i/\tau_{i+1}=\rho_*^{-\gamma}\Omega_i,
\end{equation}
where $\Omega_i$ denotes the total of the probability of all the $i$-th layer movements, as well as $\Omega_{i+1}$ for the $(i+1)$-th layer. The expression of $\Omega_i$ is:
\begin{equation}
\Omega_i = \int_0^{\Delta x_{max}}p_i(\Delta x)d(\Delta x),
\end{equation}
where $\Delta x_{max}=2\rho^{i-1}$ denotes the maximum possible length of the movements in the $i$-th layer, and $p_i(\Delta x)$ expresses the probability that an $i$-th layer movements with displacement $\Delta x$.
From Eqs. (A1) and (A2), we immediately have:
\begin{equation}
\int_0^{2\rho^i}p_{i+1}(\Delta x)d(\Delta x) = \rho_*^{-\gamma}\int_0^{2\rho^{i-1}}p_i(\Delta x)d(\Delta x).
\end{equation}

Considering $\overline{\Delta x}_{i+1} = \rho_* \overline{\Delta x}_i$, and assuming that $p_i(\Delta x)$ and $p_{i+1}$ have similar forms, the $p_i(\Delta x)$ and $p_{i+1}(\Delta x)$ will have following relationship:
\begin{equation}
p_{i+1}(\Delta x) = Cp_i(\frac{\Delta x}{\rho_*})
\end{equation}
where the constant $C$ can be approximately obtained by Eqs (A3) and (A4):
\begin{equation}
C \approx \rho_*^{-(1+\gamma)}
\end{equation}

The probability of the displacement with the relative length $\Delta x$ is the total of $p(\Delta x)$ for different layers:
\begin{equation}
P_d(\Delta x) = \sum_{i\leq j}p_i(\Delta x)
\end{equation}
where $j$ is the maximum layer number that satisfies $\Delta x\leq2\rho^{j-1}$, since the maximum possible length of the $j$-th layer movements is $2\rho^{j-1}$.

Assuming the peak of $p_j(\Delta x)$ is much closer to the averaged value $\overline{\Delta x}_j$, and considering Eqs. (A4), we have:
\begin{equation}
p_i(\overline{\Delta x}_j)|_{i<j} < \rho_*^{(1+\gamma)}p_j(\overline{\Delta x}_j)
\end{equation}
We thus have:
\begin{equation}
p_j(\overline{\Delta x}_j)<P_d(\overline{\Delta x}_j)  < \frac{1-\rho_*^{j(1+ \gamma)}}{1-\rho_*^{(1+ \gamma)}}p_j(\overline{\Delta x}_j)
\end{equation}
$P_d(\overline{\Delta x}_j)$ therefore is approximately proportional to $p_j(\overline{\Delta x}_j)$. Combining with Eq. (A4), and $\overline{\Delta x}_j = \rho_*^{j-1}$, we can get:
\begin{equation}
P_d(\overline{\Delta x}_j) \approx \rho_*^{-(1+\gamma)}P_d(\frac{\overline{\Delta x}_j}{\rho_*})
\end{equation}
We therefore obtain $P_d(\overline{\Delta x}_j) \propto \overline{\Delta x}^{-\alpha} _j$,  where the power-law exponent $\alpha = \gamma + 1$.

$P_d(\overline{\Delta x})$ for the $\overline{\Delta x}$ for different layers construct the basic power-law-like form of the displacement distribution $P_d(\Delta x)$, therefore $P_d(\Delta x)$ generally obeys a power law with the exponent $\alpha = 1+ \gamma$, even though it has slight fluctuation in the condition with small $\rho$. The analytical result is in agreement with the numerical simulations shown in Fig. \ref{Px}.
Here the exponent $\alpha$ only depends on $\gamma$, nevertheless, it is in the condition with $0 <\rho_*<1$, indicating that the cascading process in the model is the basic requirement in the origin of the power-law-like displacement distribution.


\section{Analysis for visitation frequency and the number of visited locations in the situation without cascading effect}

An extreme situation of the model without the cascading effect ($\rho* = 0$). In this case, all the movements and locations are in a same layer, and  Eq. (2) is the only rule that drives the model. The evolution of the visitation frequency distribution $P'_v(k')$ can be written directly:
\begin{equation}
kP'_v(k')=(k'-1)P'_v(k'-1),
\end{equation}
Notice here $k'$ represents the total times of visitation of a location from the initial time $t = 0$.
We obtain:
\begin{equation}
P'_v(k') \propto k'^{-1}.
\end{equation}
Considering the initial process from $t = 0$ to $t = t_0$, for $t>t_0$, we have: $P'_v(k_0+k) \propto (k_0+k)^{-1}$ and $P'_v(k_0) \propto k_0^{-1}$, where $k_0$ and $k$ respectively denote the total times of visitation of a location before and after $t_0$. So we immediately have $P'_v(k) \propto k^{-1}$, which is the visitation frequency distribution in this extreme situation.

For the growth of the number of visited locations $S'(t)$, in this extreme situation, according to the definition of $S'(t)$ and $P'_v(k)$, we have
\begin{equation}
S'(t) = \sum_k P'_v(k),
\end{equation}
and
\begin{equation}
t =  \sum_k[kP'_v(k)].
\end{equation}
Substituting $P'_v(k) \propto k^{-1}$ into the two equations above, we can easily obtain that $S'(t) \propto \ln t$.

In brief, removing the cascading effect, the visitation frequency distribution of the model will obey a power law with exponent $-1$, and the growth of the number of visited locations will show a logarithmic relationship to time.





\begin{thebibliography}{ref1}

\bibitem{bro} D. Brockmann, L. Hufnagel, T. Geisel, The scaling laws of human travel, Nature 439, 462 (2006).
\bibitem{gon} M. C. Gonz\'alez, C. A. Hidalgo, A. L. Barab\'asi. Understanding individual human mobility patterns. Nature 453, 779 (2008).
\bibitem{Rhee} I. Rhee, M. Shin, S. Hong, K. Lee, S. Chong, On the Levy-walk nature of human mobility: Do humans walk like monkeys? IEEE Conference on Computer Communications 2008, pp. 924.
\bibitem{Lee} K. Lee, S. Hong, S. J. Kim, I. Rhee, S. Chong, SLAW: A New Mobility Model for Human Walks. IEEE Conference on Computer Communications 2009, pp. 855.
\bibitem{Jiang2009} B. Jiang, J. Yin, S. Zhao, Characterizing the human mobility pattern in a large street network, Phys. Rev. E \textbf{80}, 021136 (2009).
\bibitem{Song2010b} C. Song, Koren T, Wang P, et al. Modelling the scaling properties of human mobility. Nat. Phys. 6, 818  (2010).
\bibitem{Jia} T. Jia, B. Jiang, K. Carling, M. Bolin and Y. Ban, An empirical study on human mobility and its agent-based modeling, J. Stat. Mech. P11024(2012).
\bibitem{Szell} M. Szell, R. Sinatra, G. Petri, S. Thurner, V. Latora, Understanding mobility in a social petri dish. Sci. Rep. 2, 457 (2012).

\bibitem{Song2010} C. Song, Z. Qu, N. Blumm, A.-L. Barab\'{a}si. Limits of Predictability in Human Mobility. Science 327, 1018 (2010).

\bibitem{Belik} V. Belik, T. Geisel, D. Brockmann, Natural human mobility patterns and spatial spread of infectious diseases, Phys. Rev. X 1, 011001 (2011).
\bibitem{Balcan} D. Balcan, A. Vespignani. Phase transitions in contagion processes mediated by recurrent mobility patterns. Nat. Phys. 7, 581 (2011).
\bibitem{WangL} L. Wang, X. Li, Y.-Q. Zhang, Y. Zhang, K. Zhang, Evolution of scaling emergence in large-scale spatial epidemic spreading. PLoS ONE 6, e21197 (2011).
\bibitem{Ni} S. Ni, W. Weng. Impact of travel patterns on epidemic dynamics in heterogeneous spatial metapopulation networks. Phys. Rev. E 79, 016111 (2009).
\bibitem{Wang2009} P. Wang, M. C. Gonz\'{a}lez, C. A. Hidalgo, and A. -L. Barab\'{a}si, Science \textbf{324}, 1071 (2009).
\bibitem{Zhao} Z.-D Zhao, Y. Liu, and M. Tang, Epidemic variability in hierarchical geographical networks with human activity patterns, Chaos 22, 023150 (2012).
\bibitem{Horner} M. W. Horner, M. E. S. O'Kelly, Embedding economies of scale concepts for hub networks design. J. Transp. Geogr. 9, 255 (2001).

\bibitem{Petro} S. Petrovskii, A. Mashanova, V. A. A. Jansen, Variation in individual walking behavior creates the impression of a L\'evy flight, Proc. Natl. Acad. Sci. USA 108, 8704 (2011).
\bibitem{Yan2} X. -Y. Yan, X. -P. Han, B. -H. Wang, T. Zhou, Diversity of Individual Mobility Patterns, arXiv:1211.2874.
\bibitem{Baz1} A. Bazzani, B. Giorgini, S. Rambaldi, R.Gallotti, L.Giovannini, Statistical laws in urban mobility from microscopic GPS data in the area of Florence, J. Stat. Mech. P05001 (2010).
\bibitem{Liang} X. Liang, X. Zheng, W. Lv, T. Zhu, K. Xu, The scaling of human mobility by taxis is exponential. Physica A 391, 2135 (2012).
\bibitem{Noulas} A. Noulas, S. Scellato, R. Lambiotte, M. Pontil, C. Mascolo, Tale of Many Cities: Universal Patterns in Human Urban Mobility, PloS ONE 7, e37027 (2012).
\bibitem{Peng} C. Peng, X. Jin, K.-C. Wong, M. Shi, P. Li\`o, Collective Human Mobility Pattern from Taxi Trips in Urban Area, PloS ONE 7, e34487(2012)

\bibitem{Simini} F. Simini, M. C. Gonz\'alez, A. Maritan, A.-L. Barab\'asi, A universal model for mobility and migration patterns, Nature 484, 96 (2012).
\bibitem{Vene} D. Veneziano, M. C. Gonz\'alez, Trip Length Distribution Under Multiplicative Spatial Models of Supply and Demand: Theory and Sensitivity Analysis, arxiv: 1101.3719.
\bibitem{Baz2} R. Gallotti, A. Bazzani, S. Rambaldi, Towards a statistical physics of human mobility. Int. J. Mod. Phys. C 23, 2150061 (2012).

\bibitem{Han} X. -P. Han, Q. Hao, B. -H. Wang, T. Zhou, Origin of the scaling law in human mobility: Hierarchy of traffic systems. Phys. Rev. E 83, 036117 (2011).
\bibitem{Yan1} X. -Y. Yan, X. -P. Han, T. Zhou, B. -H. Wang, Exact Solution of Gyration Radius of Individual's Trajectory for a Simplified Human Regular Mobility Model, Chin. Phys. Lett. 28, 120506 (2011).

\bibitem{XWWang} X. -W. Wang, X. -P. Han, B. -H. Wang, Autocorrelation and scaling laws in human mobility, arXiv:1303.5844.
\bibitem{Malmgren} R. D. Malmgren, D. B. Stouffer, A. E. Motter, and L. A. N. Amaral, A Poissonian explanation for heavy tails in e-mail communication, Proc. Natl. Acad. Sci. USA. 105, 18153 (2008).
\bibitem{Rybski2009} D. Rybski, S. V. Buldyrev, S. Havlin, F. Liljeros, and H. A. Makse, Scaling laws of human interaction activity, Proc. Natl. Acad. Sci. USA. 106, 12640 (2009).
\bibitem{Rybski2012} D. Rybski, S. V. Buldyrev, S. Havlin, F. Liljeros, and H. A. Makse, Communication activity in a social network: relation between long-term correlations and inter-event clustering, Sci. Rep. 2, 560 (2012).

\bibitem{bart1} F. Bartumeus, F. Peters, S. Pueyo, C. Marras\'{e}, J. Catalan, Proc. Natl. Acad. Sci. U.S.A. \textbf{100}, 12771 (2003);
\bibitem{Ramos} G. Ramos-Fern\'{a}ndez, J. L. Mateos, O. Miramontes, G. Cocho, H. Larralde, B. Ayala-Orozco, Behav. Ecol. Sociobiol. \textbf{55}, 223 (2004);
\bibitem{Sims} D. W. Sims, E. J. Southall, N. E. Humphries, G. C. Hays, C. J. A. Bradshaw, J. W. Pitchford, A. James, M. Z. Ahmed, A. S. Brierley, M. A. Hindell, D. Morritt., M. K. Musyl, D. Righton, E. L. C. Shepard, V. J. Wearmouth, R. P. Wilson, M. J. Witt, J. D. Metcalfe, Nature \textbf{451}, 1098 (2008).


\bibitem{Vqz2006} A. V\'{a}zquez, J. G. Oliveira, Z. Dezs\"{o}, K. -I. Goh, I. Kondor, and A. -L. Barab\'{a}si, Phys. Rev. E \textbf{73}, 036127 (2006).

\bibitem{Domenico} M. D. Domenico, A. Lima, M. Musolesi, Interdependence and Predictability of Human Mobility and Social Interactions, 	arXiv:1210.2376.




\end{thebibliography}
\end{document}